\documentclass[%
 reprint,
 superscriptaddress,
 amsmath,amssymb,
 aps,longbibliography
]{revtex4-2}
\usepackage{graphicx}% Include figure files
\usepackage{caption}
\usepackage{dcolumn}% Align table columns on decimal point
\usepackage{bm}% bold math
\usepackage{gensymb}
\usepackage{physics}
\usepackage[letterpaper, margin=0.6in]{geometry}
\usepackage{comment}
\usepackage[caption=false,font=footnotesize]{subfig}
\usepackage[caption=false,font=footnotesize]{subfig}
\usepackage{xcolor}
\usepackage{float}

\usepackage{hyperref}
\usepackage[utf8]{inputenc}

\hypersetup{
    colorlinks,
    citecolor=black,
    filecolor=black,
    linkcolor=black,
    urlcolor=black
}

\begin{document}
%TC:ignore

\title{Suppression of Superconductivity and Electrostatic Side Gate Tuning in High Mobility SrTiO$_3$  Surface Electron Gas}

\author{Dickson Boahen}
\thanks{These authors contributed equally to this work.}
\affiliation{Department of Physics, University of Cincinnati, Cincinnati, OH 45219, USA}

\author{Sushant Padhye}
\thanks{These authors contributed equally to this work.}
\affiliation{Department of Electrical and Computer Engineering, University of Cincinnati, Cincinnati, OH 45219, USA}

\author{Gayan De Silva}
\affiliation{Department of Physics, University of Cincinnati, Cincinnati, OH 45219, USA}

\author{Eshanvi Rao}
\affiliation{Department of Physics, University of Cincinnati, Cincinnati, OH 45219, USA}

\author{Evgeny Mikheev}
\affiliation{Department of Physics, University of Cincinnati, Cincinnati, OH 45219, USA}

\captionsetup[figure]{labelfont={normal},labelformat={default},labelsep=period,name={FIG.},justification=raggedright,singlelinecheck=false}

\begin{abstract}

We report on the fabrication and characterization of patterned high-mobility two-dimensional electron gases (2DEG) formed on SrTiO$_3$ (STO) substrate surfaces by hydrogen plasma exposure. The resulting devices consistently showed high electron mobilities up to 7400 cm$^2$/(V$\cdot$s). A large range of electron density was systematically explored by controlled aging of the sample between cooldowns, including the expected range for the STO 2DEG superconducting dome. No superconducting transition was observed down to the base temperature of approximately 10 mK. This suggests  suppression of superconductivity in high mobility quasi-two-dimensional SrTiO$_3$ electron gas, likely linked to vertical confinement and electronic orbital rearrangement. We systematically explored electrostatic gate modulation in this 2DEG system and its scaling with electron density and side gate geometry. In contrast with our initial expectation, we observed an improvement of achievable total modulation for larger side gate to channel separation. At low electron density, stochastic channel pinch-off events were observed, creating quasi-ballistic constrictions with irregular conductance quantization. This epitaxy-free and high mobility oxide material platform offers a promising new route towards patterning quantum devices.

\noindent \small $^*$ These authors contributed equally to this work.

\noindent \textbf{Corresponding authors:} Evgeny Mikheev, email: mikheev@ucmail.uc.edu

\end{abstract}
\maketitle

% \section{Introduction}
%TC:endignore

\medskip
%\subsection*{Main text}

\noindent\textbf{Introduction}

\noindent Strontium Titanate (STO) is an exciting material host for low-dimensional electron systems. It can combine unusually dilute superconductivity \cite{Gastiasoro20, Lin13}, that is electrostatically tunable \cite{Joshua12, Caviglia08}, an exceptionally high low-temperature dielectric constant \cite{Neville1972, Yang2022,Eggli23}, Rashba-type spin-orbit coupling \cite{Caviglia2010}, and micron-range electron mean free path \cite{Mikheev23}.

This system has a unique promise for realizing monolithic single crystal mesoscopic devices. Several exciting nanofabrication approaches emerged in recent years: SrTiO$_3$/LaAlO$_3$ side gating \cite{Prawiroatmodjo17, Thierschmann18, Jouan20} and hard masking \cite{Monteiro17, Stornaiuolo17}, biased AFM tip writing \cite{Annadi18}, side gate-masked ionic liquid gating \cite{Mikheev23, Mikheev21, Gallagher14}. There is an apparent dichotomy in STO 2DEG transport and nanodevice results: distinct approaches achieve either (A) clean ballistic and/or quantum transport with  superconductivity absent \cite{Annadi18, Jouan20, Mikheev23} or heavily suppressed \cite{Gallagher15, Olsen25,Caviglia10} or (B) quasi-ballistic superconducting transport with electron mean free path below 100 nm and thus significant interference from mesoscopic disorder \cite{Prawiroatmodjo16, Monteiro17, Thierschmann18, Chen18, Mikheev21,Jouan22}. In this work we systematically explore a new promising approach for patterning a high mobility STO 2DEG and demonstrate that superconductivity is absent within the millikelvin temperature regime accessible in a standard dilution refrigerator. This result emphasizes the open questions of why STO superconductivity and clean transport appear to be in competition, and how to reach the desirable regime of their coexistence.

The STO 2DEGs in this work are defined by hydrogen plasma exposure (HPE), previously demonstrated in \cite{04_takashi_wiley, 22_yadav_arxiv}. This fabrication technique is remarkably cost effective: it uses commercial SrTiO$_3$ single crystals and does not require any technically intricate thin film growth. We demonstrate here that it is compatible with e-beam lithography patterning on the few micron scale and has potential for further device feature size down-scaling towards the mesoscopic and quantum regimes.

The timescale of gradual 2DEG aging is readily adjustable between  months (in ambient storage) and hours (on a standard hot plate). This allowed for efficient and systematic exploration of a wide electron density range in search of superconductivity.

In parallel, we characterized electrostatic side gating and its scaling with channel and gate geometry to assess viability of this platform for mesoscopic oxide device development (for example quantum constrictions, junctions, wires, dots). In line with many other STO 2DEG gating works \cite{Bell09,Mikheev23}, we observe a side gate effect dominated by mobility and confinement tuning rather than capacitive charge modulation. At low electron densities in our narrowest 2DEG channels (5 $\mu$m lithographic width), we observed several instances of channel pinch-off into a quasi-ballistic constriction. Our systematic side gating characterization signals a viable, HPE-based route towards the controlled mesoscopic regime in nanoscale channels with long gate-channel gaps that mitigate leakage.

\medskip 
\noindent\textbf{Experimental Methods}

\begin{figure*}[htbp]
% \centering
\includegraphics[width=\textwidth]{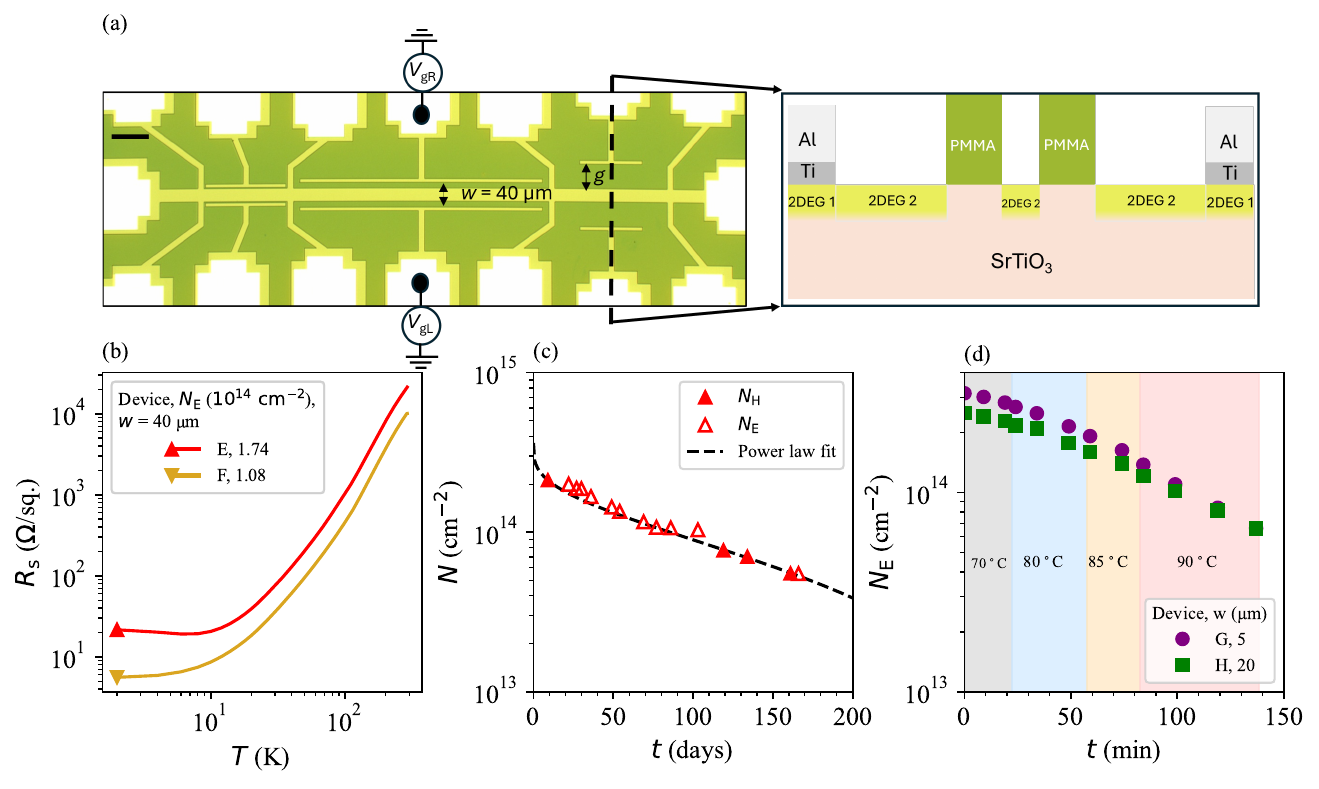}
\caption{\label{mainfig1} (a) Optical image of the device E at 5X magnification. The dashed line indicates the location of the cross-sectional schematic shown on the right. Scale bar is 100 $\mathrm{\mu}$m (b) Sheet resistance as a function of temperature for two devices measured in a pulse tube cryostat. (c) Aging of the 2DEG in nitrogen desiccator storage. Gradual decrease in electron density with time after hydrogen plasma exposure follows a power law ($N_\mathrm{H}(t=0) - At^{\alpha}$). $N_\mathrm{H}$ represents the measured Hall electron density at $\approx$10 mK and $N_\mathrm{E}$  is the estimated room temperature carrier density using a mobility value of 5.5 cm$^2$/(V$\cdot$s) \cite{Mikheev15}. (d) Accelerated 2DEG aging by mild heating of the sample. Shaded region transitions indicate hot plate temperature changes.}
\end{figure*}

\noindent Side-gated Hall bar devices (Fig. \ref{mainfig1}(a)) were fabricated in two E-beam lithography (EBL) steps, combining commercial 5x5x0.5 mm SrTiO$_3$ (001) single crystals (Nat 20 Scientific), hydrogen plasma exposure and metal lift-off. The first EBL step defines the metallic bonding pads. This resist pattern is exposed to hydrogen plasma before proceeding to a lift-off process for Ti/Al (10/200 nm) deposited by E-beam evaporation. This creates the first 2DEG region under the metallic bond pad (``2DEG1" in Fig. \ref{mainfig1}(a)), intended to decrease metal-channel contact resistance. The second EBL step defines the Hall bar 2DEG channel and side gate contacts (``2DEG2" in Fig. \ref{mainfig1}(a)), followed up by a second HPE step. Both HPE steps are performed in a Nordson RIE-1701 (see supplementary section S1 for details). For additional Van der Pauw sample testing, Ti/Al contacts were deposited in the corners and followed up with a single HPE step.

This study contained a series of devices labeled A through H with channel widths $w$ between 5 $\mu$m and 40 $\mu$m and gate-channel gaps $g$ between 5 $\mu$m and 80 $\mu$m (see table 1 in supplementary section S1 for details). Device testing combined dilution refrigerator cooldowns for milliKelvin transport and gating with pulse-tube cryostat cooldowns for wide temperature dependence of metallic transport. Gradual 2DEG aging was monitored by measuring its sheet resistance $R_\mathrm{s}$ at room temperature, which allowed for estimating electron density as $N_\mathrm{E} = (\mu e R_\mathrm{s})^{-1}$ under the assumption of a constant electron mobility $\mu=$ 5.5 cm$^2$/(V$\cdot$s) (as observed in \cite{Mikheev15})

Fig. \ref{mainfig1}(c) shows 2DEG aging of device ``E" that was stored at ambient temperature in a nitrogen desiccator in between 4 dilution refrigerator cooldowns and 1 pulse-tube cryocooler cooldown over a period of 6 months after device fabrication (Fig. \ref{mainfig1}(c)). The trend in Hall density $N_\mathrm{H}$ measured at milliKelvin temperatures matches the electron density $N_\mathrm{E}$ estimated from room temperature measurements, confirming the validity of using $N_\mathrm{E}$ as a proxy for $N_\mathrm{H}$. This aging process can be substantially accelerated (while maintaining control) by mildly heating the sample up to 70-90 \textdegree C on a hot plate in air (Fig. \ref{mainfig1}(d)). Controlled monitoring of $N_\mathrm{E}$ allowed us to explore the electron density range between initial values of $2-3 \times 10^{14}$cm$^{-2}$ and approximately $5 \times 10^{13}$cm$^{-2}$, the point where we typically saw significant non-linearity in the metal-2DEG contact resistance.  

The default hypothesis for 2DEG formation and aging mechanism is formation of oxygen vacancies (which are electron donors in STO) near the crystal surface in the reducing environment of hydrogen plasma \cite{22_yadav_arxiv}. Gradual reoxidation eliminates oxygen vacancies and reduces electron density. Within this work we cannot eliminate a possible role of hydrogen incorporation into the STO lattice, as documented for  other STO-based processes \cite{Bi10,Takeuchi20}.

\medskip 

\noindent\textbf{Suppression of superconductivity}

\begin{figure*}
    %\centering
    \includegraphics[width=\linewidth]{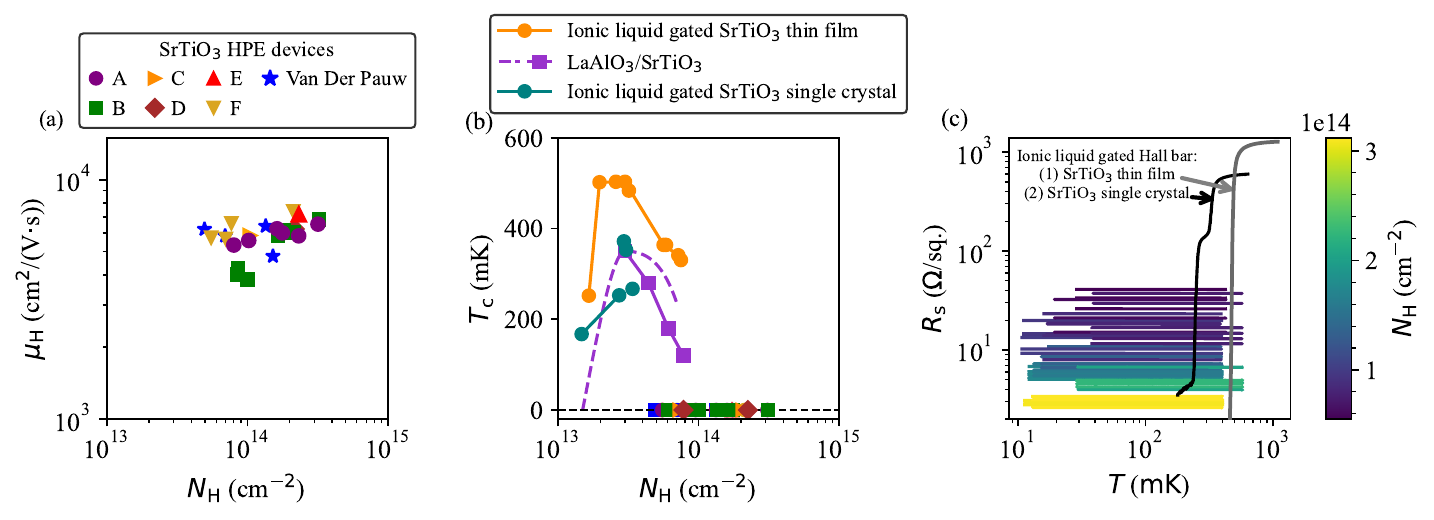}
    
    \caption{\label{mainfig2} (a) Hall mobility at milliKelvin temperatures as a function of carrier density for side-gated Hall bar devices and Van Der Pauw samples. Average value across three Hall bar voltage pairs are shown for each cooldown. (b) Electron density dependence of superconducting transition temperature in SrTiO$_3$ 2DEGs. Absence of detectable superconductivity in HPE devices in this work is marked as $T_\mathrm{c} =$ 0. The representative LaAlO$_{3}$/SrTiO$_{3}$ dome trace following \cite{Jouan22,Mikheev23} is shown as a dashed line, along with experimental data as connected symbols for LaAlO$_3$/SrTiO$_3$ heterostructures \cite{BenShalom10}, ionic liquid gated SrTiO$_3$ single crystals \cite{Mikheev21}, and ionic liquid gated SrTiO$_3$ thin films \cite{padhye26}.  (c) The measured flat temperature dependence of resistance in HPE devices A-F, with trace color mapped to Hall density. Superconducting transition traces measured in ionic liquid gated SrTiO$_3$ single crystal and thin film (from \cite{padhye26}) are shown for comparison. All data in (c) and ionic liquid gated SrTiO$_3$ thin film data in (b) were measured in the same dilution refrigerator system.}
    
\end{figure*}

\noindent All Hall bar devices and Van der Pauw samples measured at low temperature showed clean electron transport. Wide temperature range measurements of sheet resistance $R_\mathrm{s}$ (Fig. \ref{mainfig1}(b)) show robust metallic transport. The residual resistance ratio is approximately 1000, indicative of high mobility. Fig. \ref{mainfig2}(a) shows Hall mobility $\mu_\mathrm{H}$ measured at base dilution refrigerator temperatures. It is approximately independent of $N_\mathrm{H}$ with scatter within the 3800-7400 cm$^2$/(V$\cdot$s) range. The corresponding electron mean free paths are within the $0.6-2$ $\mu$m range. For comparison, STO 2DEG superconductivity is typically observed in samples with $\mu_\mathrm{H}$ below $10^3$ cm$^2$/(V$\cdot$s) \cite{Joshua12,Caviglia08,Prawiroatmodjo16, Monteiro17, Thierschmann18, Chen18, Mikheev21,Jouan22}. Slightly higher Hall mobilities of order $10^4$ cm$^2$/(V$\cdot$s) have been observed in  STO heterostructures \cite{chen15, Trier16, Rubi20, Xie14} and ionic liquid gated STO 2DEGs with HfO$_x$ and hBN barrier layers \cite{Gallagher15,Mikheev23}. 

The expected location of the superconducting dome in STO 2DEGs is, broadly, within the $10^{13}$ - $10^{14}$ cm$^{-2}$ range. The optimal $T_\mathrm{c}$ of approximately 350 mK is typically found near $3 \times 10^{13}$ cm$^{-2}$ \cite{Jouan22}. In the ``underdoped" regime, superconductivity dies near  $N_\mathrm{H} = 1$ - $2\times 10^{13}$ cm$^{-2}$ due to loss of coherence between weakly coupled superconducting puddles \cite{Prawiroatmodjo16,Chen18,Singh18}. In the ``overdoped" regime, $T_\mathrm{c}$ gradually decreases \cite{Jouan22}. But to the best of our knowledge, a gradual complete suppression of $T_\mathrm{c}$ from optimum to zero in this density regime has not yet been documented.

The $5 \times 10^{13}$ cm$^{-2}$ lower bound of the electron density range that we were able to explore is well within the slightly ``overdoped" side of this dome. Robust superconductivity above 100 mK would be expected at this density in LaAlO$_3$/SrTiO$_3$ and ionic liquid gated SrTiO$_3$ surfaces. But we did not observe any signatures of superconductivity in any of our HPE-based 2DEG Hall bar devices or Van der Pauw samples. Fig. \ref{mainfig2}(c) shows flat and featureless  $R_\mathrm{s}(T)$ traces measured across all devices and samples down to the dilution refrigerator base temperature. The $T_\mathrm{c}$ of these samples is labeled as zero in Fig. \ref{mainfig2}(b). Absence of meaningful resistance decrease with AC current excitation down to 1 nA was checked at base temperature. We emphasize that robust STO 2DEG superconductivity was contemporaneously measured on the same experimental setup. For comparison, Fig. \ref{mainfig2}(c) includes traces of superconducting transitions in ionic liquid gated SrTiO$_3$ single crystal and epitaxial SrTiO$_3$ thin film \cite{padhye26}.  The measurement lines in the dilution refrigerator are filtered using a commercial multi-stage series of RC and LC low-pass filters (QDevil Qfilters) mounted at the mixing chamber stage with additional RC filters integrated onto the Qboard sample holder.

\begin{figure*}[htbp]
\centering
\includegraphics[width=7in]{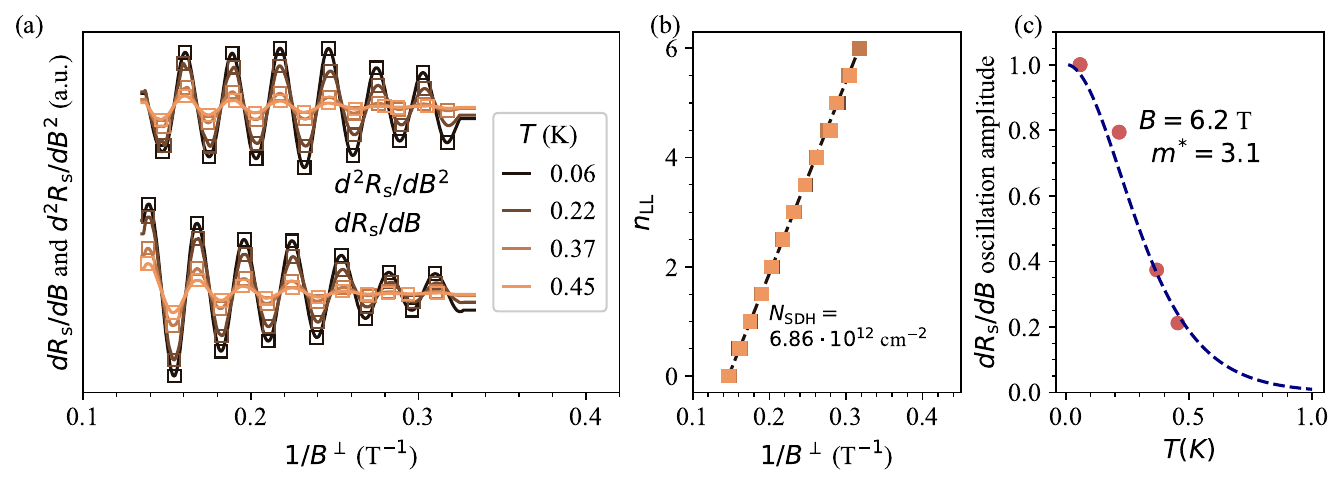}
\caption{\label{figure_sdh} (a) Shubnikov-de Haas Oscillations resistance oscillations in out-of-plane magnetic field and temperature shown in the first and second field derivatives. Markers indicate the maxima and minima. (b) Peaks and dips in $d^2R_{\mathrm{S}}/dB^2$ were used to assign Landau level indices $n_{\mathrm{LL}}$.  The dashed line shows the linear fit to electron density at the lowest temperature (60 mK). (c) Temperature dependence of the amplitude of the oscillations in the first derivative at $B=6.2$ T. The dashed line shows the fit to the Lifshitz-Kosevich model with $m^*=3.1$.}
\end{figure*}

The mechanism for the observed suppression of superconductivity is an open question, but the vertical confinement profile is an obvious important factor to consider. It is widely acknowledged \cite{Khalsa12,Chen16,Chen18} that most SrTiO$_3$ 2DEGs are not strictly two-dimensional, but rather quasi-two-dimensional with a sheet electron density of $10^{13}$-$10^{14}$ cm$^{-2}$ unevenly distributed across at least several unit cell layers. Vertical extent estimates vary widely: 1-10 nm from numerical modeling \cite{Khalsa12}, 0.65 $\mu$m from Hall mobility comparison between HPE 2DEGs and bulk Nb-doped STO \cite{04_takashi_wiley}, 4 nm from hydrogen ion penetration depth in STO \cite{22_yadav_arxiv}, 6-7 nm from quantum point contact subband magnetic field evolution in high mobility ionic liquid gated 2DEGs \cite{Mikheev23}, 5-10 nm in low mobility superconducting STO \cite{padhye26}. High mobility is typically associated with deeper confinement profiles that minimize surface disorder scattering \cite{Chen16}. A conservative lower bound for volume electron density in our device is $5\times10^{17}$ cm$^{-3}$, assuming $5\times10^{13}$ cm$^{-2}$ distributed over 1 $\mu$m thickness. The superconducting dome in uniformly-doped bulk SrTiO$_3$ extends into this range with $T_\mathrm{c} \approx$ 100 mK \cite{Lin13,Gastiasoro20}. So, a simple explanation of volume density being reduced below the onset of superconductivity is unlikely to apply. 

Another important consideration is the redistribution of orbital occupation between shallow and low-mass $d_{xy}$  and deeper lying, heavy-mass $d_{xz/yz}$ states. Typically, $d_{xy}$ states are filled first, and the Lifshitz transition to multi-band occupation including $d_{xz/yz}$ states occurs near optimal $T_\mathrm{c}$ \cite{Joshua12}. The debate on which band flavor favors superconductivity remains open, with several recent works on back-gating and numerical modeling in LaAlO$_3$/SrTiO$_3$ 2DEGs indicating enhancement of $T_\mathrm{c}$ with increased $d_{xz/yz}$ population \cite{Wjcik24, Jouan22}. In possible contradiction, band order inversion towards single doubly-degenerate $d_{xz/yz}$ band filling has been observed in high-mobility 2DEG systems with suppressed superconductivity: SrTiO$_3$/$\gamma$-Alumina \cite{Chikina21, Olsen25} and ionic liquid-gated SrTiO$_3$ with HfO$_x$ barrier layers \cite{Mikheev23}. 

The HPE-based STO 2DEGs in this work appear to also belong to the $d_{xz/yz}$ filling archetype, as suggested by the relatively high electron mass from quantum oscillations.  Fig. \ref{figure_sdh}(a) shows oscillations measured at $N_{\mathrm{H}}=5.1\times10^{13}$ cm$^{-2}$, presented as first and second derivatives with field. Identifying oscillation extrema in $d^2R/dB^2$ as Landau levels, their frequency $f_{\mathrm{SDH}}$ in $1/B$ converts to an electron density of $6.86\times10^{12}$ cm$^{-2}$ with  $N_{\mathrm{SDH}}=f_{\mathrm{SDH}}v^{-1}_{\mathrm{s}}eh^{-1}$ and spin degeneracy $v_{\mathrm{s}}=1$. The factor of $\approx10$ discrepancy between $N_{\mathrm{SDH}}$ and $N_{\mathrm{H}}$ is a ubiquitous feature of STO 2DEGs  \cite{Caviglia10,Jalan10,Rubi20,Mikheev23}. Temperature dependence of the oscillation amplitude at $B=6.2$ T (Fig. \ref{figure_sdh}(c)) was used to extract the effective electron mass from a fit to the Lifshitz-Kosevich model: $\delta R_{\mathrm{xx}}\sim \alpha T/\mathrm{sinh}(\alpha T)$, where $\alpha=2\pi^2\mathrm{k_B}/\mathrm{\hbar}\omega_{\mathrm{c}}$ is the thermal suppression factor and the cyclotron frequency is $\omega_{\mathrm{c}}=\mathrm{e}B/m_{\mathrm{e}}^*$. The resulting high value for the effective mass $m_{\mathrm{e}}^*=3.1m_{\mathrm{e}}$ suggests occupation of the heavy-mass $d_{xz/yz}$ states, in line with \cite{Mikheev23,Chikina21}.

\medskip 
\noindent\textbf{Side-Gate Modulation}

\begin{figure*}[htbp]
% \centering
\includegraphics[width=\textwidth]{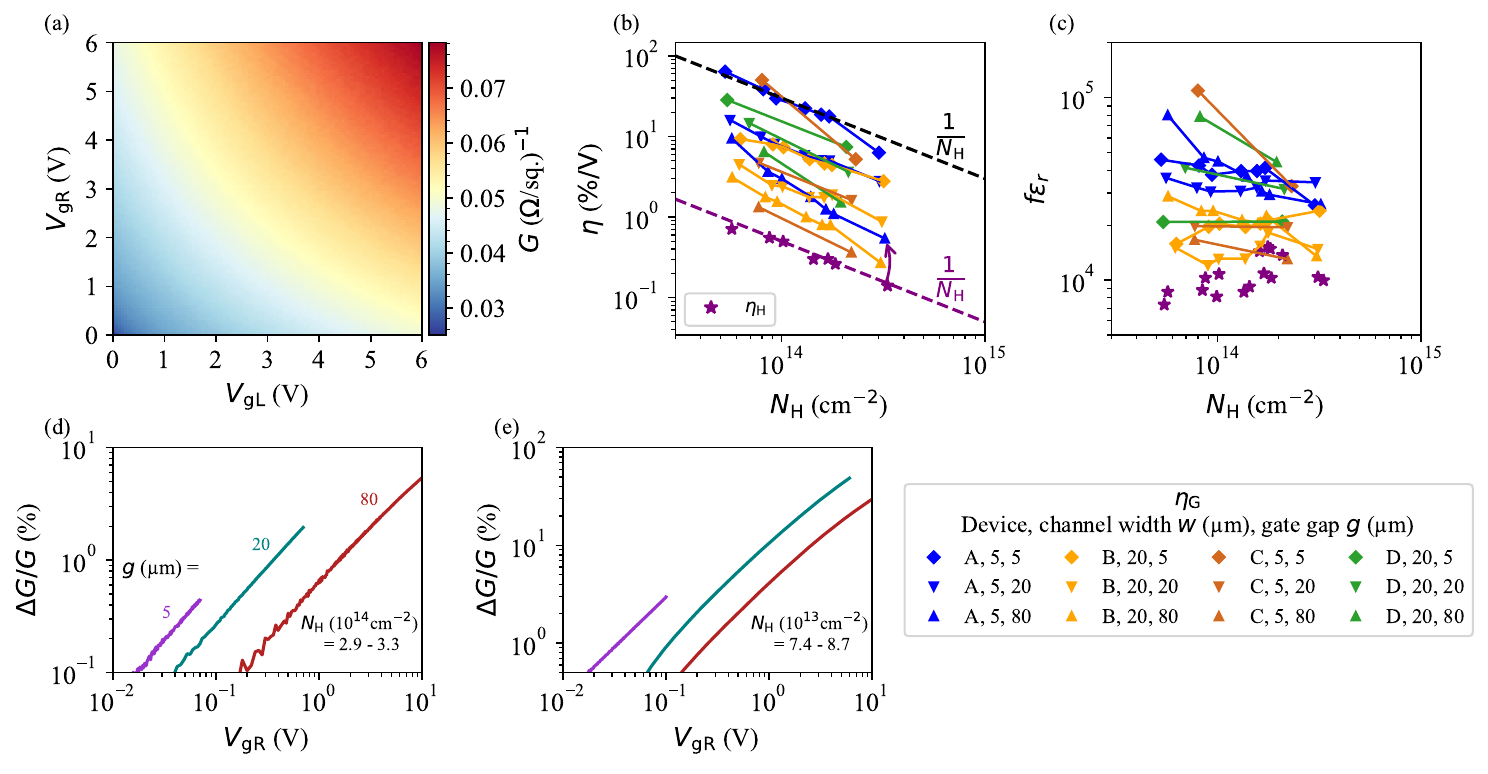}
\caption{\label{mainfig4} (a) Symmetric modulation of channel conductance by two adjacent side gates in device A, $w=$ 5 $\mu$m, $g=$ 20 $\mu$m. (b) Side gate modulation efficiency in conductance ($\eta_G$) and Hall density ($\eta_N$) as a function of electron density for devices A-D. Dashed lines highlight $N_\mathrm{H}^{-1}$ scaling. (c) Equivalent effective dielectric constant with lumped geometric factor. For conductance modulation, mobility modulation by is also lumped into $f\varepsilon_r$. (d,e) Conductance modulation vs side gate voltage in two representative cooldowns of device A, highlighting a higher total modulation in remote gates due to lower leakage and despite lower gating efficiency.   }
\end{figure*}

\noindent In comparison with most substrates for mesoscopic and quantum devices, side gating from remote gate contacts is unusually efficient in SrTiO$_3$ at low temperatures due to its large dielectric constant of order $10^4$ from quantum paraeletric behavior. Our Hall bar devices included 3 pairs of side gates with distinct distances $g$ to the 2DEG channel between 5 and 80 $\mu$m. The T-shaped gate contacts were defined within the same lithography step and HPE as the Hall bar Channel. This 2DEG-insulator-2DEG gate design was intended to minimize the parasitic low permittivity series capacitance at SrTiO$_3$/metal interfaces \cite{Mikheev14}. For all cooldowns, we systematically characterized the response of channel conductance to the left and right side gate voltages ($V_{\mathrm{gL}}$ and $V_{\mathrm{gR}}$, respectively). As shown in the two-gate sweep in Fig. \ref{mainfig4}(a), conductance modulation was typically symmetric. This indicates reproducible gate capacitance at the lithographically identical left and right side gates. The initial gate sweeps had substantial hysteresis, similar to \cite{Biscaras14,Mikheev23}, but the subsequent stable sweeps were nearly free of hysteresis (see supplementary section S4). No meaningful signatures of incipient superconductivity were detected.

To inform future device design, we quantified side gate efficiency $\eta_\mathrm{G} = G^{-1} (\partial G/\partial V_\mathrm{g})$ in a stable sweep. For a Drude conductance $G = eN_\mathrm{H}\mu_\mathrm{H}$, electrostatic charge modulation by a gate capacitance $C_\mathrm{g}$ should give $\eta_G = C_\mathrm{g}/N_\mathrm{H}$. This $N_\mathrm{H}^{-1}$ trend broadly holds in Fig. \ref{mainfig4}(b) when considering each individual device. Some traces show stronger $N_\mathrm{H}$ dependence (but weaker than $N_\mathrm{H}^{-2}$).

The Hall density modulation $\eta_\mathrm{N} = N_\mathrm{H}^{-1} (\partial N_\mathrm{H}/\partial V_\mathrm{g})$ was systematically monitored for one device next to a selected side gate ($w=$ 5 $\mu$m, $g=$ 80 $\mu$m). It is shown in Fig. \ref{mainfig4}(b) as purple stars and follows the $N_\mathrm{H}^{-1}$ scaling expected for a fixed $C_\mathrm{g}$. The corresponding $\eta_\mathrm{G}$ in the adjacent channel section is indicated by an arrow in Fig. \ref{mainfig4}(b) and is consistently higher than $\eta_\mathrm{N}$ by a factor of 5-10. The slightly longer channel-gate gap in the $\eta_\mathrm{N}$ measurement accounts for a much smaller factor of $\sqrt{2}$. This discrepancy is consistent with gate modulation of the 2DEG vertical confinement profile, leading to a deeper 2DEG with reduced surface scattering and increasing Hall mobility at higher gate voltage \cite{Chen16}.

To quantify geometric dependence on $w$ and $g$, we invoke a minimal model approximating side gate as a parallel plate capacitor. Gate capacitance is assumed to $C_\mathrm{g}=f\varepsilon_\mathrm{0}\varepsilon_\mathrm{r}/z$, with $f$ a geometric factor of order unity \cite{Eggli23} and $z=g+w/2$. The effective dielectric constant $f\varepsilon_\mathrm{r}$ estimated from $C_\mathrm{g}=\eta_\mathrm{N}N_\mathrm{H}$ is 8,000 - 15,000, slightly below the typically measured  $\varepsilon_\mathrm{r}=$ 20,000 - 30,000 in STO at cryogenic temperatures \cite{Neville1972, Yang2022,Eggli23}. Estimating $f\varepsilon_\mathrm{r}$ from resistive modulation with $C_\mathrm{g}=\eta_\mathrm{G} N_\mathrm{H}$ gives a range from 10,000 to 50,000, with several outliers near $10^5$. The increased value of this estimate is again a reflection of side gate effect on 2DEG confinement and electron mobility, and not of an abnormally high dielectric permittivity.

For future device design rules, minimizing $w$ and $g$ maximizes gate effect per volt. However, minimizing $w$ only is a better guideline for maximizing the total achievable modulation. We observed a sharp increase in gate leakage threshold for shorter $g$ (see supplementary section S4). Fig. \ref{mainfig4}(d,e) show conductance modulation against side gate voltage in device A for two representative cooldowns at high and low $N_\mathrm{H}$. The $g=$ 5 $\mu$m gate follows the steepest linear slope, but due to early leakage onset, its total achievable modulation is substantially lower than the $g=$ 20 and 80 $\mu$m gates. The most remote $g=$ 80 $\mu$m achieves the highest total modulation since it can be pushed to higher gate voltage.

\medskip
\noindent\textbf{Pinch-off and Conductance Quantization}

\begin{figure*}[htbp]
\centering
\includegraphics[width=7in]{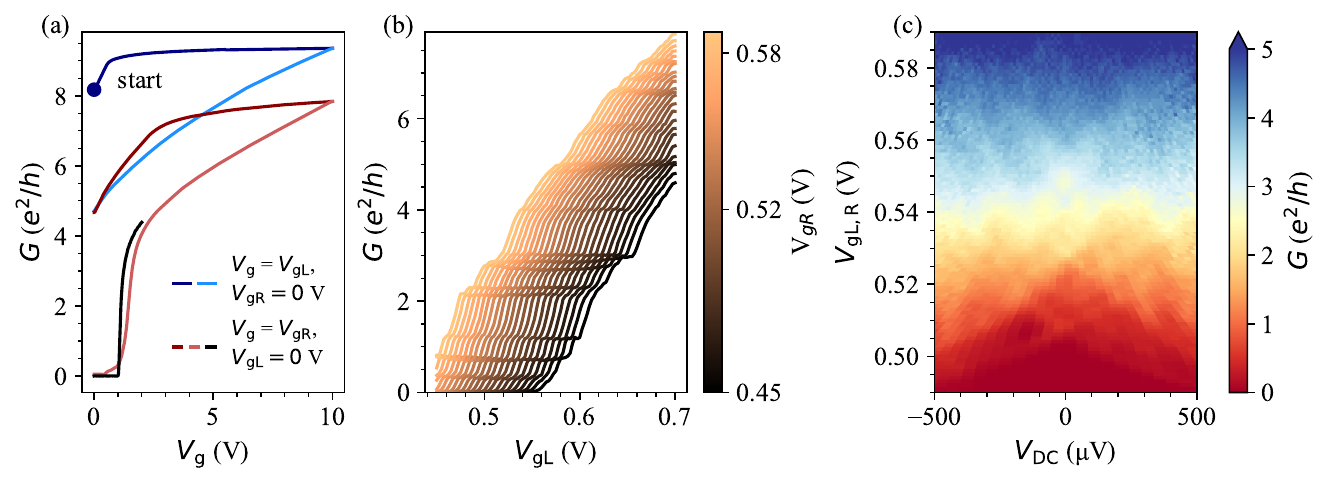}
\caption{\label{figure_pinchoff} (a) Initial side gate voltage ramping on the $g=$ 20 $\mu$m section of device A at 5.5 $\times$ 10$^{13}$ cm$^{-2}$. $V_\mathrm{gL}$ is ramped to 10 V and back to zero first, followed by $V_\mathrm{gR}$. The black trace is a stable sweep with $V_\mathrm{gR}$. (b) Conductance was measured by ramping $V_{\mathrm{gL}}$ at various constant $V_{\mathrm{gR}}$ values, showing repeatable steps that slowly evolve with disorder potential tuning by $V_{\mathrm{gR}}$ . (c) DC bias spectroscopy showing quasi-ballistic subbands. $V_{\mathrm{DC}}$ is the measured DC voltage drop across the relevant channel section. Both gates were swept at equal gate voltage $V_\mathrm{gL}=V_\mathrm{gR}=V_\mathrm{gL,R}$. Series resistance of 4 k$\Omega$ was subtracted in (b) and (c).}
\end{figure*}

During the cooldowns at lower carrier densities, we observed sharp resistance increases during initial hysteretic side gate sweeps in one of the 5 $\mu$m-wide devices. Fig. \ref{figure_pinchoff}(a) shows the initial ramping procedure for 20-$\mu$m gap side gates at 8.4 $\times$ 10$^{13}$ cm$^{-2}$. $V_{\mathrm{gL}}$ and $V_{\mathrm{gR}}$ were individually swept from 0 V to 10 V and back with the other side gate grounded. Channel conductance in subsequent stable sweeps is pinched off at small positive side gate voltages (Fig. \ref{figure_pinchoff}(b)). Supplementary Fig. S8  shows similar pinch-off events in two distinct device A sections by 20 and 80-$\mu$m gap side gates at 5.5 $\times$ 10$^{13}$ cm$^{-2}$.

The sudden onset of channel pinch-off can be attributed to channel inhomogeneity, which has been extensively documented in SrTiO$_3$ 2DEGs and correlated with the tetragonal domain structure and their polarity at the nanoscale \cite{13_kalisky_natmat, 21_persky_ncomms, 16_ma_physrevlett, 17_frenkel_natmat}. The relevant channel section in Fig. \ref{figure_pinchoff}(c) is 840 $\mu$m long, offering a substantial 2DEG area for an unintentional weak spot to emerge from statistical fluctuation. This interpretation is supported by Fig. \ref{figure_pinchoff}(c), which shows quasi-ballistic transport consistent with a local constriction. 
 
We observed repeatable step features in conductance - side gate voltage traces near $G$ of order conductance quantum. The ideal ballistic quantum point contact picture gives conductance plateaus at integer multiples of $G=2e^2/h$ in a spin degenerate 2DEG \cite{Buttiker90}. In our case, a much larger number of steps is present. This is commonly observed in intentional quasi-ballistic constrictions of size comparable to the electron mean free path in SrTiO$_3$ 2DEGs \cite{Mikheev21} and other semiconductors \cite{06_jiang_natnanotech, 19_mittag_physrevb}. Exact assignment of step features to conductance quantum multiples is not trivial due to uncertainty on series resistance. Subtracting a constant 4 k$\Omega$ series resistor in Fig. \ref{figure_pinchoff}b() results in an irregular step sequence feature clustering  near 1, 2, 3 and 5 $e^2/h$ with faint step near 4 and 6 $e^2/h$. Plots without series resistance subtraction are shown in supplementary section S5.

The quasi-ballistic nature of this constriction is confirmed by DC bias spectroscopy in Fig. \ref{figure_pinchoff}(c). The step features are revealed as zero bias crossings between subbands of electrons moving in opposite directions through the constriction. The $100-200$ $\mu$eV width of the resulting diamonds is very similar to both the vertical and lateral confinement length scales in approximately 40-nm wide SrTiO$_3$ 2DEG constrictions in \cite{Mikheev23}.

\medskip 
\noindent\textbf{Conclusion}

\noindent This work demonstrates a straightforward and cost effective approach to patterning side-gated Hall bar devices in high-mobility STO 2DEGs. This process relies solely on hydrogen plasma exposure,  standard lithographic techniques and commercially available single crystals, with no involvement of advanced epitaxial growth techniques.

Absence of detectable superconductivity within our systematic search window in electron density  reinforces the sparsely documented trend on absent or suppressed superconductivity in clean STO 2DEGs with micron-scale electron mean free path \cite{Mikheev23,Gallagher15, Olsen25,Caviglia10}. It is an open question whether this suppression is primarily driven by disorder reduction, rearrangement of vertical 2DEG confinement or orbital occupation. Practically, a key future challenge is to tune the STO 2DEG towards the crossover regime between the firmly established limits of dirty superconducting and clean normal state 2DEGs. This compromise regime is desirable for fabrication of ballistic superconducting nanodevices and for understanding the evolution of superconducting length scales in the quasi-2D clean limit of STO.

\subsection*{Acknowledgments}

The authors acknowledge Huma Yusuf and Ron Flenniken for assistance with device fabrication, and Yashar Komijani for insightful discussions. This work used the Advanced Materials Characterization Center and the Mantei Center Cleanroom at University of Cincinnati. This work was primarily supported by the Office of Naval Research through award N00014-24-1-2079. Early development of the hydrogen plasma exposure techniques was supported by the National Science Foundation through award DMR-2328826-000.

\subsection*{Data Availability}

Raw data, python notebooks used during original data collection, and python notebooks for generation of manuscript figures are available at https://doi.org/10.5281/zenodo.19745634

\medskip 

\subsection*{References}

\medskip
\renewcommand{\bibsection}{\section*{}}
\bibliography{references.bib}

% \documentclass[%
%  reprint,
%  superscriptaddress,
%  amsmath,amssymb,
%  aps,longbibliography
% ]{revtex4-2}

% \usepackage{graphicx}% Include figure files
% \usepackage{caption}
% \usepackage{dcolumn}% Align table columns on decimal point
% \usepackage{bm}% bold math
% \usepackage{gensymb}
% \usepackage{bbold}
% \usepackage{physics}
% \usepackage{comment}
% \usepackage{float}
% \usepackage{placeins}

% \usepackage[
% backend=biber,
% style=science,
% ]{biblatex}
% \addbibresource{references.bib} %Imports bibliography file

% \usepackage{hyperref}
% \usepackage[utf8]{inputenc}

\hypersetup{
    colorlinks,
    citecolor=black,
    filecolor=black,
    linkcolor=black,
    urlcolor=black
}

%%%%%%%%%%%%%%%%%%%
% \preprint{APS/123-QED}
% \begin{document}

\onecolumngrid
\captionsetup[figure]{labelfont={bf},labelformat={default},labelsep=period,name={Fig.}}
\setcounter{figure}{0}
\renewcommand{\thefigure}{S\arabic{figure}}
\setcounter{page}{1}
\renewcommand{\thepage}{S\arabic{page}}
\setcounter{section}{0}
\renewcommand{\thesection}{S\arabic{section}}
\setcounter{equation}{0}
\renewcommand{\theequation}{S\arabic{equation}}

\section*{Supplementary material for ‘‘Suppression of Superconductivity and Electrostatic Side Gate Tuning in High Mobility SrTiO$_3$  Surface Electron Gas’’}

% \renewcommand{\citenumfont}[1]{S#1}
% \renewcommand{\bibnumfmt}[1]{[S#1]}

% \tableofcontents

% \clearpage

\section{Device Fabrication}

\label{SMsection1}
\noindent Commercial 5x5x0.5 mm, (001)-oriented SrTiO$_3$ (STO) single crystal substrates (Nat 20 Scientific)  were first cleaned by ultrasonication in acetone for one minute, followed by a 3-minute rinse in isopropanol (IPA), and then blow-dried with a nitrogen gun. An insulating layer of PMMA A8 resist was spin-coated on the substrate at 4000 rpm to achieve a thickness of approximately one micron. The sample was baked on a hot plate at 110 \textdegree C for 10 minutes post-resist spin. To prevent surface charging during electron beam lithography (EBL), a conductive polymer, Electra E92, was spun on top of the PMMA using the same spin coating recipe. The first EBL step defined the metallic bonding pads. Nabity Nanopatterning Generator System (NPGS) connected to an FEI SCIOS scanning electron microscope was used. The sample was soaked in a distilled water for about 30 seconds to dissolve the conducting polymer coating and then transferred into a 3:1 water to IPA developer solution for 3 minutes. To ensure good ohmic contact between the metal and the STO 2DEG defined in the second lithographic step, the first resist pattern was initially exposed to hydrogen plasma in a Nordson RIE-1701 reactive ion etcher, creating locally conductive 2DEGs underneath the contacts. Devices A,B,C,D,G, and H were exposed to hydrogen plasma for 5 minutes at a pressure of 210 - 230 mTorr while devices E and F were exposed for 7.5 minutes at 250 - 300 mTorr. The RF power was maintained at 75 W throughout the process. This was followed by E-beam evaporation of  10 nm titanium (Ti) and  200 nm of aluminum (Al) on the same resist pattern. The sample was soaked in acetone overnight to lift off the metal, followed by a 3-minute IPA rinse. Ti/Al contacts for the Van der Pauw devices were deposited using a shadow mask technique and did not require liftoff.
The second lithographic step followed the same resist spinning, EBL, and development procedure to define the device channel and side gates, connecting their leads to the metallic bonding pads. Pattern development was followed by a 5 - 7 minutes exposure to hydrogen plasma (and the same process parameters as in the first step) to create the STO 2DEG channel. 

\begin{figure}[!b]
\centering
\includegraphics[width=7in]{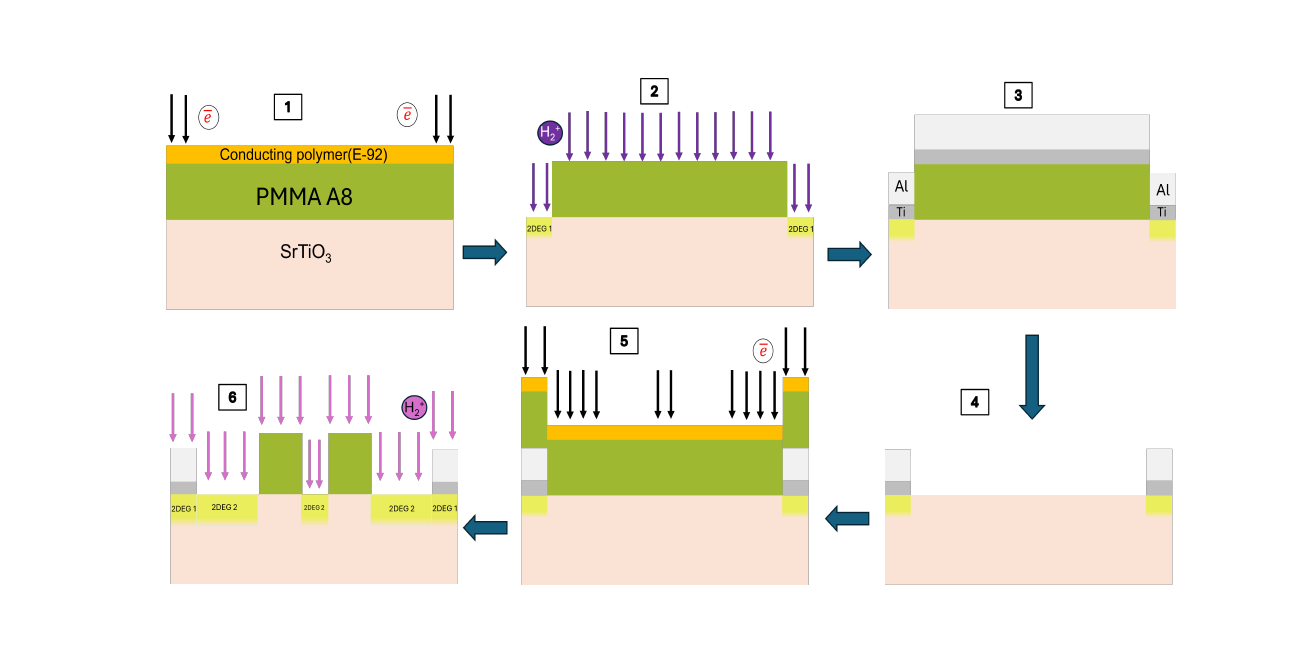}
\caption{\label{SM_devfabsteps} \textbf{Device Fabrication steps.} Cross-section illustration for  1) First E-beam lithography step, 2) first  hydrogen plasma exposure, 3) Titanium/ Aluminum evaporation, 4) lift-off, 5) Second E-beam lithography step, 6) Second hydrogen plasma exposure.  }
\end{figure}

\begin{table}
    \centering
    \begin{tabular}{|c|c|c|c|}
\hline
        Device & Channel width & Channel-gate gaps  & Cooldown \#,  \\
        &  $w$ ($\mu$m) &$g$ ($\mu$m) & measured Hall density range $N_\mathrm{H}$ (10$^{13}$ cm$^{-2}$)\\
\hline
        A & 5 & 5, 20, 80 & 1, 30--33 \\
        & & & 2, 17--18 \\
        & & & 3, 15-17 \\
        & & & 4, 13--14 \\
        & & & 5, 8.6--10.1\\
        & & & 6, 7.4--8.8\\
        & & & 7, 5.1--5.7\\
\hline
        B & 20 & 5, 20, 80 & 1, 30--32\\
        & & & 2, 16--18\\
        & & & 3, 15--16\\
        & & & 4, 13--14\\
        & & & 5, 9.5--10.7\\
        & & & 6, 8.3--9.6\\
        & & & 7, 5.7--6.7\\
\hline
        C & 5 & 5, 20, 80  & 1,18--21 \\
        & & & 2, 5.4--8.4\\
\hline
        D & 20 & 5, 20, 80  & 21--23\\
        & & & 7.4--8.0\\
\hline
        E & 40 & 5, 20, 80  & 1, 20--21\\
        & & & 2, 7.7\\
        & & & 3, 5.7--7.1\\
        & & & 4, 4.2--5.5\\
\hline
        F & 40 & 20, 60 & 1, 10.1--10.3\\
\hline
        G & 5 & 5, 20, 80  & From room temperature estimates: \\
        & & & 6.0--32\\
\hline
        H & 20 & 5, 20, 80  &  From room temperature estimates: \\
        & & & 6.4--24\\

\hline
    \end{tabular}
    \caption{Summary of fabricated device geometries and dilution refrigerator cooldowns with measured Hall density ranges across the three device sections. Device pairs A and B, C and D, G and H were fabricated on the same chip and characterized in parallel.}
    \label{SM_dev_cooldowninfo}
\end{table}

\begin{figure}[h]
\centering
\includegraphics[width=7in]{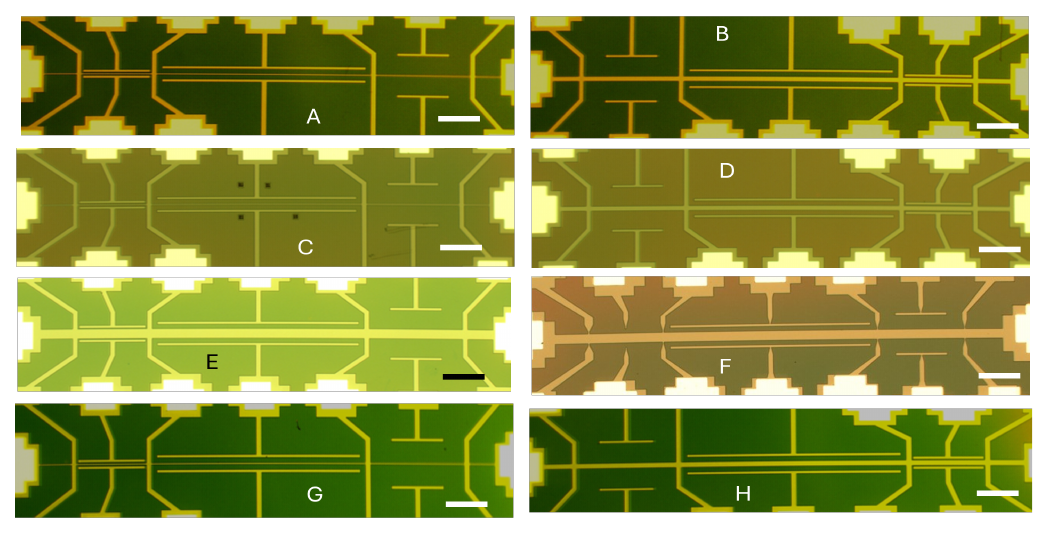}
\caption{\label{SM_fabricateddevices} \textbf{Optical images.} Images of devices ``A" through ``H" taken at 5X magnification after the second hydrogen plasma exposure. The scale bars are 200 $\mu$m wide}
\end{figure}

\section{Experimental Setup and Measurement Techniques}
\label{SMsection2}
MilliKelvin transport measurements were carried out using a Bluefors LD400 dilution refrigerator with a 9-3 Tesla 2D vector magnet. The sample is wire-bonded to a Qdevil Qboard printed circuit daughterboard and inserted into the dilution unit via a bottom-loading fast sample exchange (FSE). Following 2-terminal contact checks with a Keithley 2450 Sourcemeter, all 4-terminal measurements were carried out with four  SR860 lock-in amplifiers, three Basel Precision Instruments SP1004 voltage preamps, and one Basel Precision Instruments SP983c-IF current preamp. Gate voltage was set with either a Keithley 2450 Sourcemeter or a Qdevil QDAC-II 24-channel DAC.

For complementary wide temperature range transport characterization, a pulse tube (PT) cryostat (Lakeshore Shi-4-2) with a base temperature of $2.3$ K.  A Keithley DAQ6510 Data Acquisition System was coupled to a Keithley 7710 multiplexer for automated 4-terminal DC voltage measurement across all six voltage probe pairs in our devices. A Keithley 2400 sourcemeter was used for DC excitation sourcing.

\begin{figure}[htbp]
\centering
\includegraphics[width=4.5in]{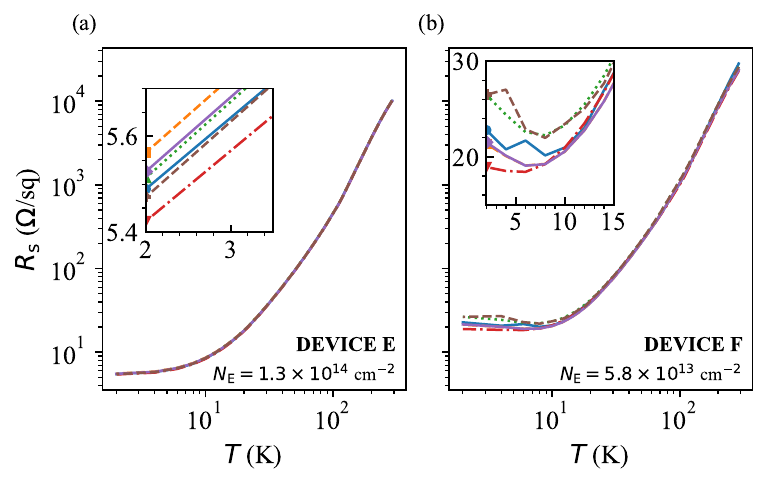}
\caption{\label{SM_RsvsT_devEF} \textbf{Representative temperature dependence of the sheet resistance} $R_\mathrm{s}$ for device E (a) and F (b), measured  in the pulse-tube cryostat. The different line styles correspond to $R_\mathrm{s}$ traces for all the pairs of voltage taps on the devices.}
\end{figure}

The temperature dependence of 2DEG sheet resistances upon cooling from room temperature to 2.3 K is shown in Fig. \ref{SM_RsvsT_devEF} for devices E and F. Traces measured across different pairs overlap very closely, confirming high macroscopic uniformity of our 2DEG along the Hall bar channel. Small deviations between traces at low temperature are consistent with mesoscopic inhomogeneity affecting disorder scattering.

\clearpage

\section{Device aging}
\label{SMsection3}

\begin{figure}[H]
\centering
\includegraphics[width=7.5in]{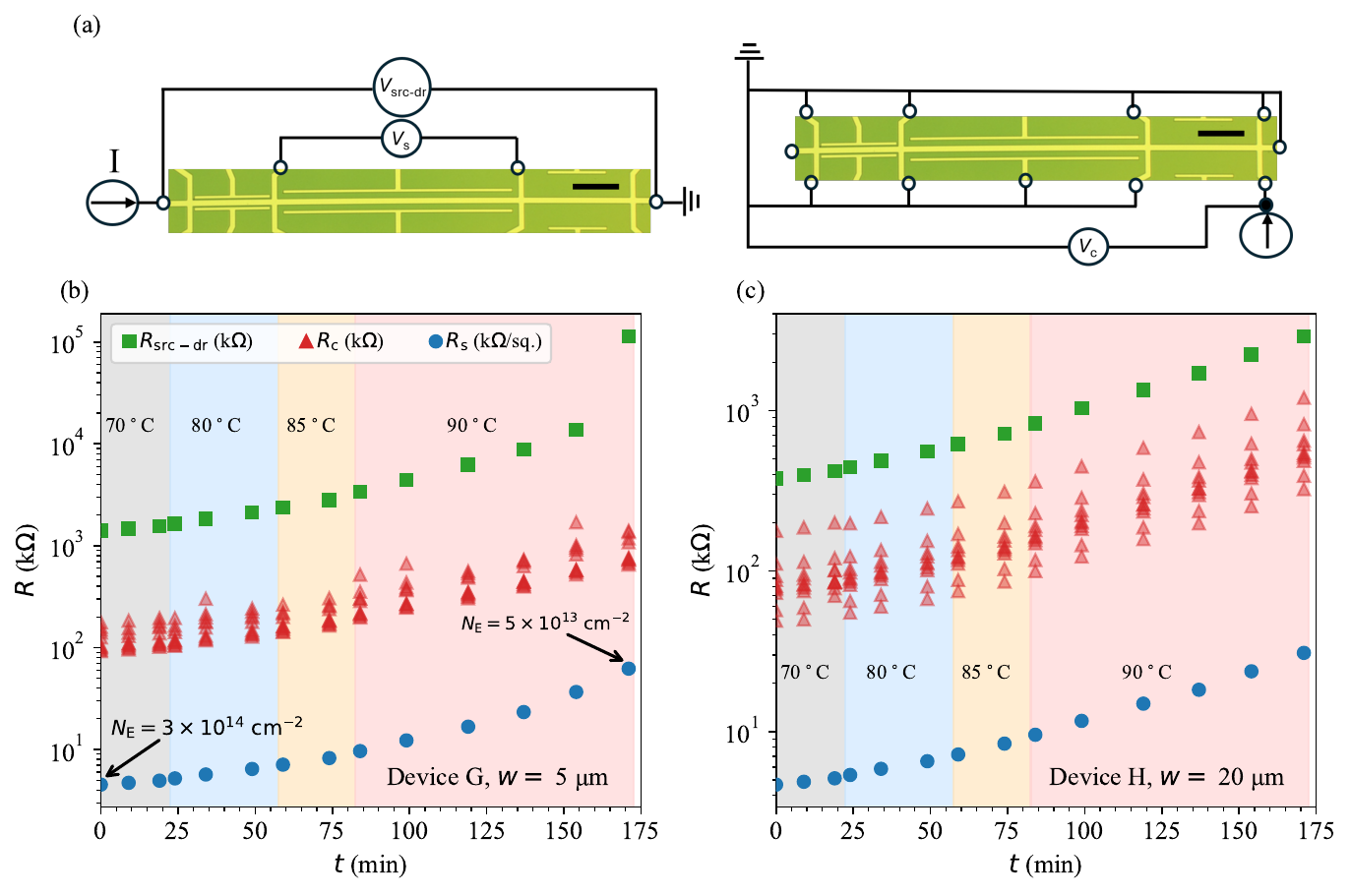}
\caption{\label{SM_resistancemeasurements} (a) Measurement configurations for room temperature Hall bar monitoring: two-terminal source-drain voltage $V_\mathrm{src-dr}$, four-terminal longitudinal voltage $V_\mathrm{s}$, and two-terminal contact voltage $V_\mathrm{c}$. (b),(c) Source-drain resistance $R_\mathrm{src-dr}$, contact resistance $R_\mathrm{c}$, and the sheet resistance $R_\mathrm{s}$ versus time on the hot plate for devices ``G" ($w=$ 5 $\mu$m) and ``H" ($w=$ 20 $\mu$m), respectively. The shaded regions indicate hotplate temperature adjustments.}
\end{figure}

Device aging can be conveniently monitored through simple four terminal resistance measurements at room temperature. It enables quick estimation of electron density reduction, and thus better informed choice of device state for dilution refrigerator cooldowns. Figure \ref{SM_resistancemeasurements} presents detailed characterization of four- and two- terminal resistance evolution in two devices (``G" and ``H") subjected to accelerated aging at slightly elevated temperatures in air. Fig. \ref{SM_resistancemeasurements}(a) illustrates the three measurement configurations: the two-terminal source-drain resistance $R_\mathrm{src-dr}$ across the Hall bar channel with all intermediate contact floated, the four-terminal sheet resistance $R_\mathrm{s}$ between two longitudinal voltage taps, and the two-terminal contact resistance $R_\mathrm{c}$ between one voltage tap contact (to which the excitation is applied) and all other contacts shorted to the ground. Their evolution with time on the hot plate is shown in Figs. \ref{SM_resistancemeasurements}(b) and \ref{SM_resistancemeasurements}(c). All contact resistances gradually increase similarly to the 2DEG sheet resistance. Our devices typically  failed close to an estimated electron density of $N_\mathrm{E}=5 \times 10^{13}$ cm$^{-2}$, either from rapid increase of contact resistance (as for $R_\mathrm{src-dr}$ in Fig. \ref{SM_resistancemeasurements}(b)) or rapid onset of strong non-linearity in contact current-voltage dependence.

\begin{figure}[H]
\centering
\includegraphics[width=7.5in]{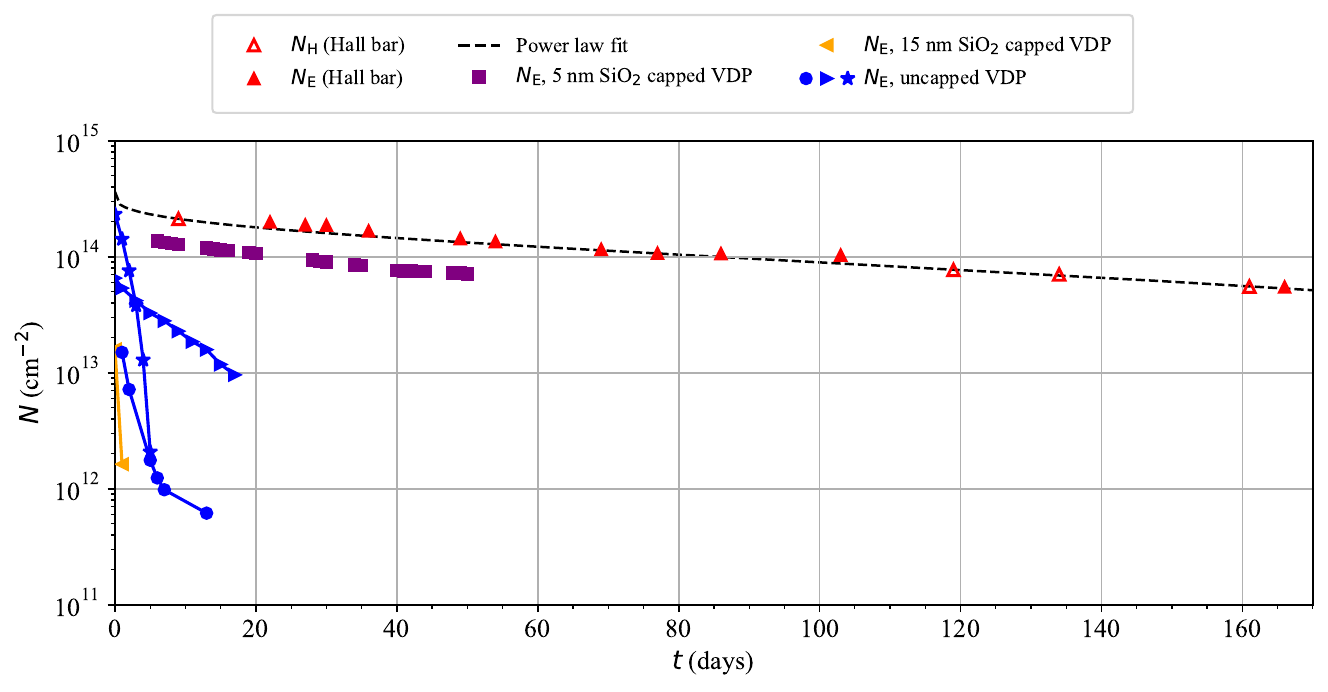}
\caption{\label{SM_agingHPEHB_VDP} Electron density as a function of time after hydrogen plasma exposure for Hall bar and VDP samples. Open red symbols indicate measured Hall density $N_\mathrm{H}$ at base temperature ($\sim$ 10 mK), while filled symbols represent carrier density $N_\mathrm{E}$ estimated from room temperature $R_\mathrm{s}$. 2DEG density for the uncapped VDP samples decreases much quicker than in Hall bar devices. But VDP sample capped with a 5nm SiO$_2$ layer shows significantly slower aging.}
\end{figure}

Fig. \ref{SM_agingHPEHB_VDP} shows supplementary data on 2DEG aging in Van der Pauw (VDW) samples. In comparison to Hall bar devices, VDP samples showed significantly faster aging at ambient temperature. Simple Van der Pauw samples with bare SrTiO$_3$ surfaces exposed to hydrogen plasma (blue symbols in Fig. \ref{SM_agingHPEHB_VDP}) typically showed measurable 2DEG conductivity for approximately a week, and often presented challenges with cryogenic freeze out of metal-2DEG contacts. The difference in aging timescales appears to be connected to 2DEG area, and we speculate that large lateral length scales facilitate gradual oxygen diffusion towards oxygen vacancies.

Fig. \ref{SM_agingHPEHB_VDP} also shows preliminary results on 2DEG capping. with sputtered silicon oxide. The tested approach involved room temperature sputtering of SiO$_2$ onto SrTiO$_3$ with Ti/Al corner contacts, followed by hydrogen plasma exposure. For a 30 nm SiO$_2$ capping layer, no measurable conductivity was observed. Reducing the SiO$_2$ thickness to 15 nm resulted in a weak 2DEG which rapidly became unmeasurable within one day (yellow symbols in Fig. \ref{SM_agingHPEHB_VDP}). Using a 5 nm SiO$_2$ capping layer (purple symbols in Fig. \ref{SM_agingHPEHB_VDP})  produced a robust 2DEG with a slow aging rate, comparable to an uncapped Hall bar device. Therefore, Hydrogen plasma is capable of penetrating a thin layer of SiO$_2$, which subsequently slows down the 2DEG aging process. This combination offers a promising route to long-term stable 2DEG devices.

\clearpage
\section{Gate voltage tuning}
\label{SMsection5}
\begin{figure}[htbp]
\centering
\includegraphics[width=7.5in]{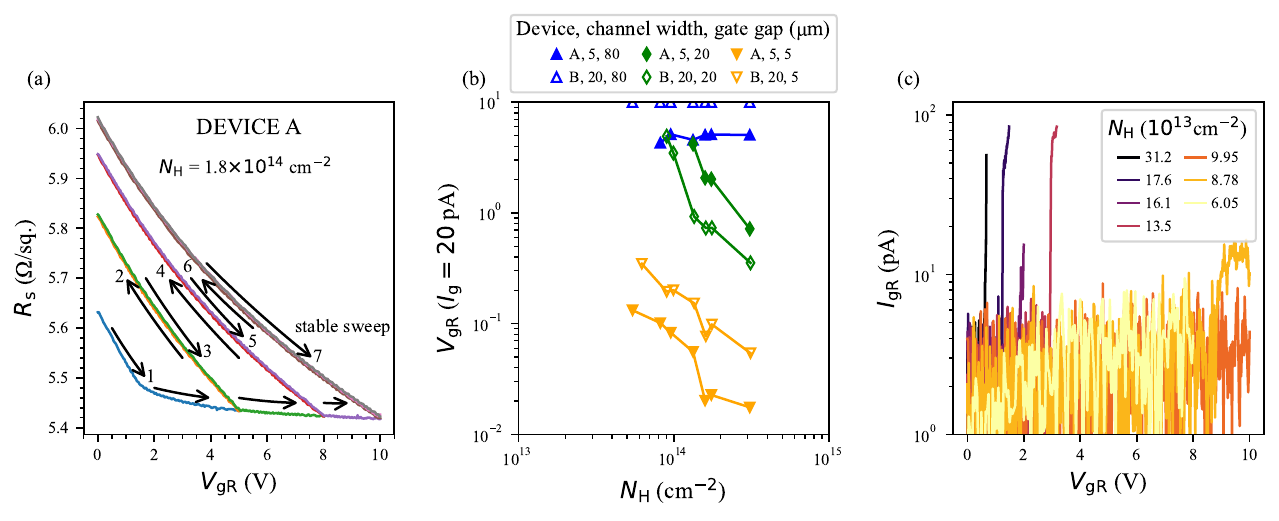}
\caption{\label{SM_gatinggraphs} \textbf{Gate voltage sweeps at base temperature and observed leakage} (a) Representative gate voltage $V_g$ sweep on the 80 $\mu$m gate. $V_\mathrm{g}$ was gradually swept up and down in several steps from 0 to 10 V. (b) Gate voltage threshold for gate leakage current above 20 pA in devices A and B. For device B, the $g=$ 80$ \mu$m gate did show leakage below 20 pA across the entire tested gate voltage range up to 10 volts. (c) Evolution of leakage current on the 20 $\mu$m gate (device B) across seven cooldowns showing a substantial improvement in gate leakage at the lowest carrier density }
\end{figure}

Initial side gate voltage sweeps on the gates were performed at base dilution refrigerator temperatures using a Keithley 2400 source meter. Fig. \ref{SM_gatinggraphs}(a) shows hysteretic behavior observed in all initial sweeps. 2DEG resistance decrease quickly saturates and returns to a higher zero gate voltage value than initially. Subsequent gate sweeps within the previously swept gate voltage range are stable and non-hysteretic.

During these initial sweeps, gate leakage current was closely monitored through the sourcemeter for sharp increases to define the upper bound of accessible gate voltage range. The threshold for gate leakage onset consistently increased as the electron density decreased with device aging. Example raw data for leakage current against gate voltage and its evolution with $N_\mathrm{H}$ are shown in Fig. \ref{SM_gatinggraphs}(c) for the 20-micron channel-gate gap in device A.

Fig. \ref{SM_gatinggraphs}(b) shows electron density evolution of the side gate leakage onset threshold (set at 20 pA). This threshold is consistently much lower for 5 $\mu$m than 20 and 80 $\mu$m channel-gate gaps. This leads to higher achievable modulation with remote gates, despite lower modulation per volt.

\section{Extended Data for Pinch-off and Conductance Quantization}
\label{SMsection6}

\begin{figure*}[htbp]
\centering
\includegraphics[width=7in]{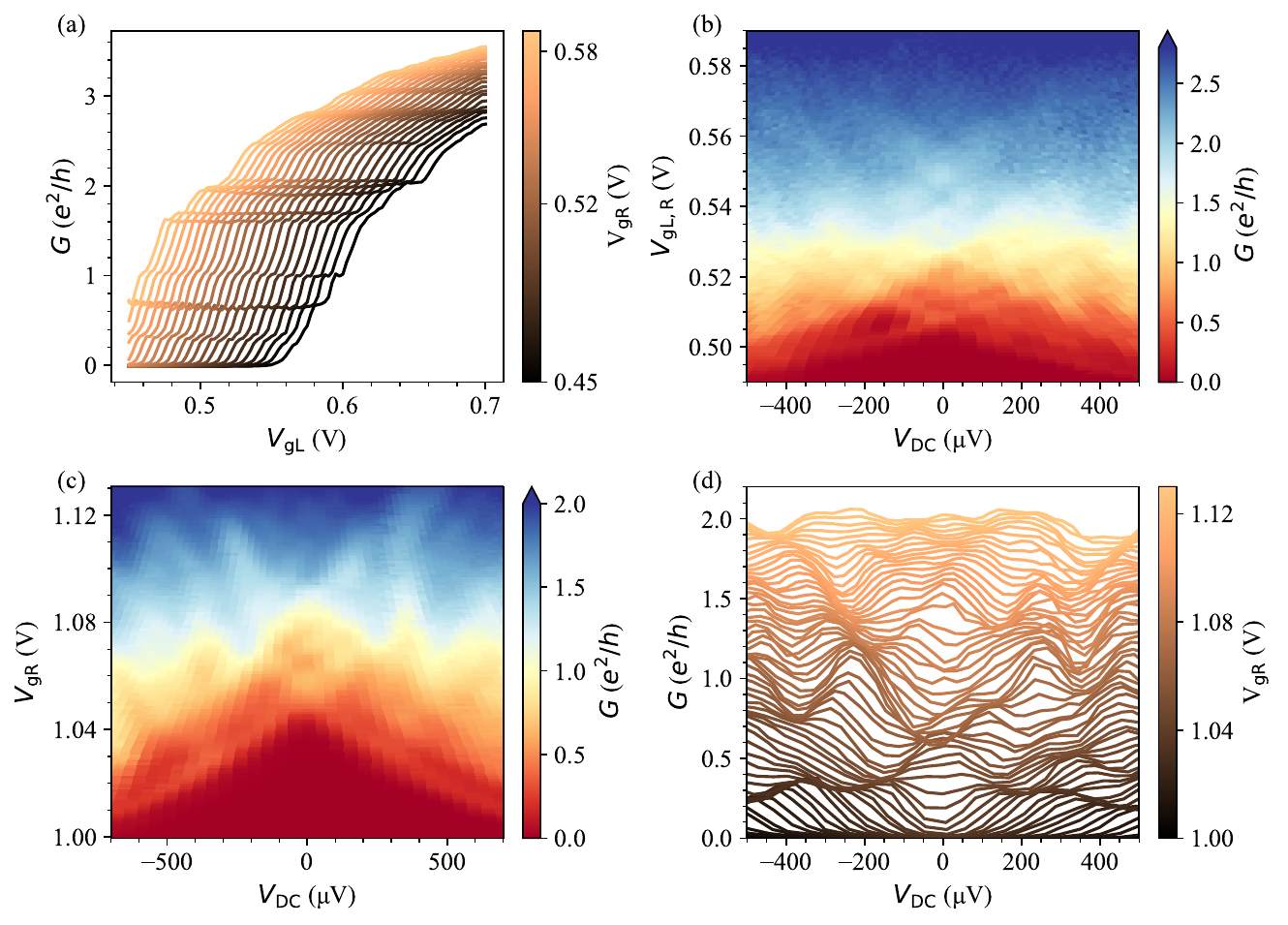}
\caption{\label{figure_supp_pinchoff_fig1} \textbf{Raw data on DC Bias Spectroscopy and Conductance Quantization} (a), (b) Same data as Figs. 5(b) and 5(c) of main text, shown without any series resistance subtraction. (c) Channel was tuned using $V_{\mathrm{gR}}$ only. A DC excitation was sourced through the channel and $V_{\mathrm{DC}}$ is the measured  DC voltage drop across the relevant channel section. (d) Same data as (c), replotted as line traces of conductance at fixed $V_{\mathrm{gR}}$.}
\end{figure*}

\begin{figure*}[htbp]
\centering
\includegraphics[width=7in]{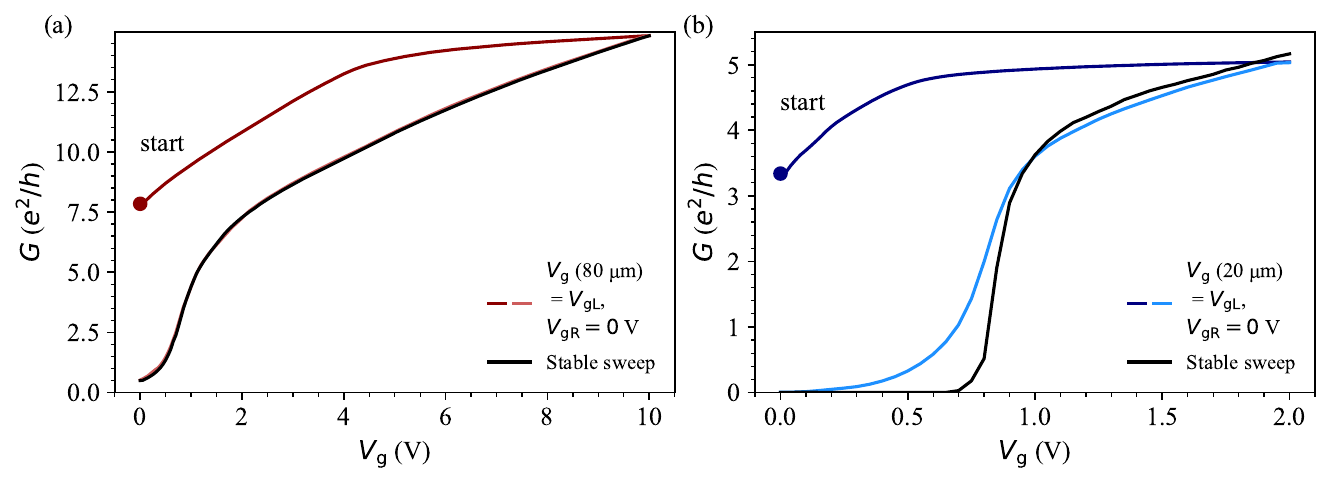}
\caption{\label{figure_supp_pinchoff_fig2} Additional channel pinch-off events in the initial $V_{\mathrm{g}}$ ramping on the (a) 80 and (b) 20 $\mu$m far side-gates respectively at $N_{\mathrm{H}}=5.5 \times 10^{13}$ cm$^{-2}$ in device A.}
\end{figure*}

Fig. \ref{figure_supp_pinchoff_fig1}(c) and \ref{figure_supp_pinchoff_fig1}(d) shows DC bias spectroscopy data at $N_{\mathrm{H}}=8.4 \times 10^{13}$ cm$^{-2}$ using only $V_{\mathrm{gR}}$  on the $g=$ 20 $\mu$m side-gate pair, with $V_{\mathrm{gL}}=0$ V (rather than a combined $V_{\mathrm{gR}}$ and $V_{\mathrm{gL}}$ sweep). This measurement was taken before the measurement shown in Fig. 5(c) in the main text. It also shows conductance trace bunching at non-integer multiples of conductance quantum, indicative of quasi-ballistic sub-band transport through a disordered constriction. 

Fig. \ref{figure_supp_pinchoff_fig2} shows the measurements at $N_{\mathrm{H}}=5.5 \times 10^{13}$ cm$^{-2}$, where the initial voltage sweeps on both the $g=$ 20 and 80 $\mu$m side gates led to sharp conductance decrease on the return trace and channel pinch-off.

\end{document}